\documentstyle[aps,prl,epsf,floats]{revtex}
\def\lsi{\raise0.3ex\hbox{$<$\kern-0.75em\raise-1.1ex\hbox{$\sim$}}}
\def\gsi{\raise0.3ex\hbox{$>$\kern-0.75em\raise-1.1ex\hbox{$\sim$}}}
\def\ha{H_1}
\def\hb{H_2}
\def\DD{{\cal D}}

\begin{document}
\twocolumn[\hsize\textwidth\columnwidth\hsize\csname
@twocolumnfalse\endcsname

\title{Electroweak Phase Transition in the MSSM: 4-Dimensional 
Lattice Simulations}
\author{F.~Csikor$^{\rm a}$, Z.~Fodor$^{\rm a}$, P.~Heged\"us$^{\rm a}$, 
A.~Jakov\'ac$^{\rm b}$, 
S.~D.~Katz$^{\rm a}$, A.~Pir\'oth$^{\rm a}$}
\address{$^{\rm a}$ Institute for Theoretical Physics, E\"otv\"os
University, H-1117, P\'azm\'any P. 1A, Budapest, Hungary}
\address{$^{\rm b}$ Department of Theoretical Physics, Technical
University, Budapest, H-1521, Budapest, Hungary
}
\date{\today}
\maketitle


\begin{abstract}\noindent
Recent lattice results have shown that there is no
Standard Model (SM) electroweak phase transition (EWPT) 
for Higgs boson masses above $\approx \!$ 72 GeV, which is below 
the present experimental limit. According to perturbation theory 
and 3-dimensional (3d) lattice simulations there could be
an EWPT in the Minimal Supersymmetric Standard Model (MSSM) that is 
strong enough for baryogenesis up to $m_h \! \approx \! 105$ GeV. 
In this letter we present the results of our large scale 4-dimensional (4d)
lattice simulations for the MSSM EWPT. We carried out infinite volume
and continuum limits  and found a transition whose strength agrees well with
perturbation theory, allowing MSSM electroweak baryogenesis 
at least up to  $m_h = 103  \pm 4$ GeV. We determined the properties of the
bubble wall that are important for a successful baryogenesis.\\

\vspace*{0.2cm}

\end{abstract}

\vspace*{0.2cm}


\vskip1.5pc]

The visible Universe is made up of matter. This statement is mainly based on
observations of the cosmic diffuse $\gamma$-ray background, which
would be larger than the present limits if boundaries between
``worlds'' and ``anti-worlds'' existed \cite{cohen98}.
The observed baryon asymmetry of the universe was eventually determined
at the EWPT \cite{KRS85}. On the one hand this phase transition was the
last instance when baryon asymmetry could have been generated, around
$T \! \approx \! 100\!-\!200$ GeV.
On the other hand at these temperatures any B+L asymmetry
could have been washed out. The possibility of baryogenesis at
the EWPT is particularly attractive, since the underlying physics
can be---and has already largely been---tested in collider experiments.

The first detailed description of the EWPT in the SM was based on
perturbative techniques \cite{4d_pert}, which resulted in ${\cal O}(100\%)$ 
corrections between different orders of the perturbative expansion
for Higgs boson masses larger than about 60 GeV. The 
dimensionally reduced 3d effective model (e.g.~\cite{3d_pert}) was also
studied perturbatively and gave
similar conclusions. Large scale numerical simulations
both on 4d and 3d lattices were needed to analyze the nature of
the transition for realistic Higgs boson masses 
\cite{4d_latt,3d_latt}. These results are in complete agreement, 
and predict \cite{3d_end,4d_end} 
an end point for the first order EWPT at Higgs boson 
mass 72.0$\pm$1.4 GeV \cite{4d_end}, above which only a 
rapid cross-over can be seen.
The present experimental lower limit of the SM Higgs boson mass is 
by several standard deviations larger than the end point value, thus any EWPT
in the SM is excluded. This also means that the SM baryogenesis in the
early Universe is ruled out.

In order to explain the observed baryon asymmetry, extended
versions of the SM are necessary. Clearly, the most attractive
possibility is the MSSM. According to perturbative predictions the
EWPT could be much stronger in the MSSM than in the SM \cite{mssm_pert},
in particular if the stop mass is smaller than the top mass
\cite{light_stop}.
At two-loop level stop-gluon graphs give a considerable
strengthening of the EWPT  
(e.g.~third and fourth paper of \cite{mssm_pert}). 
A reduced 3d version of the MSSM has recently been studied on the
lattice \cite{mssm_3d} (including
${\mathrm{SU(3)}} \! \times \! {\mathrm{SU(2)}}$
gauge fields, the right-handed stop and the ``light'' combination
of the Higgses). The results show that the EWPT can
be strong enough, i.e.\ $v/T_c \! > \! 1$, up to 
$m_h \! \approx \! 105$ GeV and $m_{\tilde t} \! \approx \! 165$ GeV
(where $m_h$ is the mass of the lightest neutral scalar and
$m_{\tilde t}$ is that of the stop squark).
The possibility of spontaneous CP violation for a successful baryogenesis 
is also addresed \cite{spont_CP}.

In this letter we study the EWPT in the MSSM on 4d lattices
and carry out infinite volume and continuum limit extrapolations.
Except for the 
U(1) sector and scalars
with small Yukawa couplings, the whole bosonic sector of the MSSM is
kept: SU(3) and SU(2) gauge bosons, two Higgs doublets, left-handed and
right-handed stops and sbottoms. As it has been done in the SM case
 \cite{4d_end},
fermions, owing to their heavy Matsubara modes, are
included perturbatively in the final result.
This work extends the 3d study \cite{mssm_3d} in two ways: 
\\
a) We use 4d lattices instead of 3d.
Note, that due to very soft modes---close
to the end point in the SM---much more CPU time is needed in 4d than in 3d. 
However, this difficulty does not appear in the MSSM because the phase 
transition is strong and the dominant correlation lengths are
not that large in units of $T_c^{-1}$. Using unimproved lattice
actions the leading corrections due 
to the finite lattice spacings are proportional to $a$ in 3d and 
only to $a^2$ in 4d. For O($a$) improvement in the 3d case cf. \cite{Moore}. 
In 4d simulations we also have direct
control over zero temperature renormalization effects.
\\
b) We include both Higgs doublets, not only the light combination. 
According to standard baryogenesis scenarios (see e.g. \cite{mssm_gen})
the generated baryon number is directly proportional to the change of
$\beta$ through the bubble wall: $\Delta \beta$. 
($\tan \beta=v_2/v_1$, where $v_{1,2}$ are the expectation
values of the two Higgses.) 

The continuum lagrangian of the above theory in standard notation reads
\begin{equation}
{\cal L}={\cal L}_g+{\cal L}_k+{\cal L}_V+{\cal L}_{sm}+{\cal L}_Y+
{\cal L}_w+{\cal L}_s.
\end{equation}
The gauge part,
$ {\cal L}_g=1/4\cdot F^{(w)}_{\mu\nu}F^{(w)\mu\nu}+
1/4\cdot F^{(s)}_{\mu\nu}F^{(s)\mu\nu} 
$
is the sum of weak and strong terms. The kinetic part is the sum of the
covariant derivative terms of the two Higgs doublets ($H_1,H_2$), 
the left-handed stop-sbottom doublet ($Q$), and the right-handed
stop, sbottom
singlets ($U,D$):
$
{\cal L}_k = 
(\DD^{(w)}_\mu \ha)^\dagger (\DD^{(w) \mu} \ha)+
(\DD^{(w)}_\mu \hb)^\dagger (\DD^{(w) \mu} \hb)+ 
(\DD^{(ws)}_\mu Q)^\dagger  (\DD^{(ws) \mu} Q)+
(\DD^{(s)}_\mu U^*)^\dagger (\DD^{(s) \mu} U^*)+ 
(\DD^{(s)}_\mu D^*)^\dagger (\DD^{(s) \mu} D^*).
$
The potential term for the Higgs fields reads
$
{\cal L}_V= m_{12}^2 [\alpha_1|\ha|^2+ \alpha_2|\hb|^2- 
(\ha^\dagger \tilde{\hb}+ 
h.c.)]+ 
g_w^2/8\cdot (|\ha|^4+ |\hb|^4- 2|\ha|^2|\hb|^2+ 4|\ha^\dagger\hb|^2),
$
for which two dimensionless mass terms are defined,
$\alpha_1=m_1^2/m_{12}^2$ 
and
$\alpha_2=m_2^2/m_{12}^2.$
One gets 
$
{\cal L}_{sm}= m_Q^2 |Q|^2+m_U^2 |U|^2+m_D^2 |D|^2
$
for the squark mass part, and 
$
{\cal L}_Y= h_t^2(|QU|^2+ |\hb|^2|U|^2+ |Q^\dagger\tilde{\hb}|^2)
$
for the dominant Yukawa part. 
The quartic parts
containing the squark fields read
$
{\cal L}_w= g_w^2/8\cdot [2 \{Q \}^4- |Q|^4+
4|\ha^\dagger Q|^2+ 4|\hb^\dagger Q|^2- 
2|\ha|^2 |Q|^2- 2 |\hb|^2 |Q|^2] 
$ 
and
$
{\cal L}_s= g_s^2/8\cdot \left[3 \{Q \}^4- |Q|^4+ 2|U|^4+ 2|D|^4-
6|QU|^2 \right. 
- 6|QD|^2+ 6|U^\dagger D|^2+ 2|Q|^2|U|^2 
\left. + 2|Q|^2|D|^2- 2 |U|^2|D|^2 \right],
$
where 
$
\{Q\}^4=Q^*_{i\alpha}Q^*_{j\beta}Q_{i\beta}Q_{j\alpha}.
$
The scalar trilinear couplings have been omitted for simplicity. It is straightforward to obtain
the lattice action, for which we used the standard
Wilson plaquette, hopping and site terms.

\begin{figure}[t]
\centerline{\epsfxsize=6.5cm\epsfbox{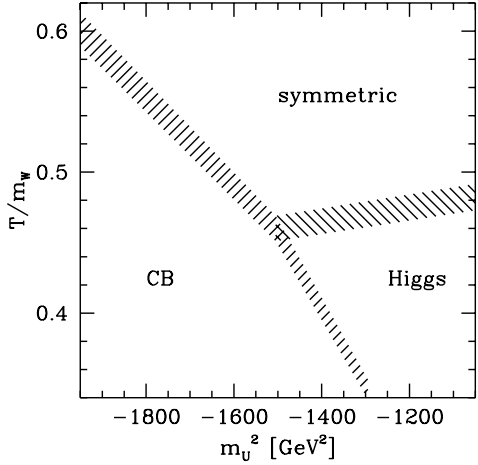}}
\caption[a]{The phase diagram of the bosonic theory obtained by lattice 
simulations.}\label{phase}
\end{figure}

The parameter space of the above Lagrangian is many-dimensional.  
We analyze the effect of the strong sector on the EWPT 
by using three specific sets of parameters.
In one case the strong coupling has its physical value, whereas
in the two other cases it is somewhat larger and smaller.
The experimental values are taken for the weak and Yukawa couplings, 
and $\tan \beta=6$
is used. For the bare soft breaking masses our choice is $m_{Q,D}=250$ GeV, 
$m_U$=0 GeV. Lattice renormalization effects on these 
masses will be discussed later.

The simulation techniques are similar to those of the SU(2)-Higgs model
\cite{4d_latt} (overrelaxation and heatbath algorithms are used
for each scalar and gauge field); some new methods
will be published elsewhere \cite{mssm_prd}. The 
analysis is based on finite temperature 
simulations (in which the temporal extension of the lattice $L_t$ is much
smaller than the spatial extensions $L_{x,y,z}$), and zero temperature
ones (with $L_t \! \approx \! L_{x,y,z}$). 
For a given $L_t$, we fix all parameters of the Lagrangian 
except $\alpha_2$. We tune $\alpha_2$ to 
the transition point, $\alpha_{2c}$, 
where we determine the jump of the Higgs field, the
shape of the bubble wall, and the change of $\beta$ through the phase boundary. 
Using $\alpha_{2c}$ and the parameter set of the finite temperature case,
we perform $T=0$ simulations and determine the masses (Higgses and W) 
and couplings (weak and strong) there. 
Extrapolations to the continuum limit and to infinite volumes are
based on simulations at temporal extensions $L_t=2,3,4,5$ and at various
lattice volumes for each $L_t$, respectively. Approaching the 
continuum limit, we move on an approximate line of constant physics
(LCP), on which the renormalized quantities (masses and couplings) are
almost constant, but the lattice spacing approaches zero.  
Our theory is  bosonic, therefore the leading corrections
due to finite lattice spacings are expected to be proportional to 
$a^2$. This lattice spacing dependence is 
assumed for physical quantities in $a\longrightarrow 0$ extrapolations.

We compare our simulation results with perturbation theory.
We used one-loop perturbation theory without applying high
temperature expansion (HTE). A specific feature was a careful treatment 
of finite renormalization effects, by 
taking into account all renormalization corrections and adjusting them 
to match the measured $T=0$ spectrum  
\cite{mssm_prd}. We studied also the effect of the
dominant $T\neq 0$ two-loop diagram (``setting-sun'' stop-gluon
graphs, cf.\ fifth ref.\ of \cite{mssm_pert}), but only in the HTE. We
observed less dramatic enhancement of the strength of the phase
transition due to two-loop effects than in \cite{mssm_pert}. 
Since the infrared behavior of the setting-sun graphs is not
understood, we use the one-loop technique with the $T=0$ scheme defined 
above.
This type of one-loop perturbation theory is also applied to correct
the measured data to some fixed LCP quantities, which are
defined as the averages of results at different lattice spacings, (i.e.\ our 
reference point, for which the most important quantity is the lightest 
Higgs mass, $m_h \approx$45 GeV).
  
Fig.~\ref{phase}. shows the phase diagram in the
$m_U^2$--T plane. One identifies three phases. The phase on the left
(large negative $m_U^2$ and small stop mass) is the
``color-breaking'' (CB) phase. The phase in the upper right part is the
``symmetric'' phase, whereas the ``Higgs'' phase can be found in the
lower right part.  The line separating the symmetric and Higgs phases
is obtained from $L_t=3$ simulations, whereas the lines between these
phases and the CB one are determined by keeping the lattice
spacing fixed while increasing and decreasing the temperature by
changing $L_t$ to 2 and 4, respectively. The shaded regions indicate
the uncertainty in the critical temperatures.  The phase transition to
the CB phase is observed to be much stronger than that
between the symmetric and Higgs phases. The qualitative features of
this picture are in complete agreement with perturbative and 3d
lattice results \cite{mssm_pert,light_stop,mssm_3d}; however, our choice of
parameters does not correspond to a two-stage symmetric-Higgs phase 
transition. In this 
two-stage scenario there is a phase transition from the symmetric to
the CB phase at some $T_1$ and another phase
transition occurs at $T_2<T_1$ from the CB to the Higgs phase.  
It has been argued \cite{cline99} that in the early universe no
two-stage phase transition took place, therefore we do not study this
possibility and the features of the CB phase any further.

\begin{figure}[t]
\centerline{\epsfxsize=6.5cm\epsfbox{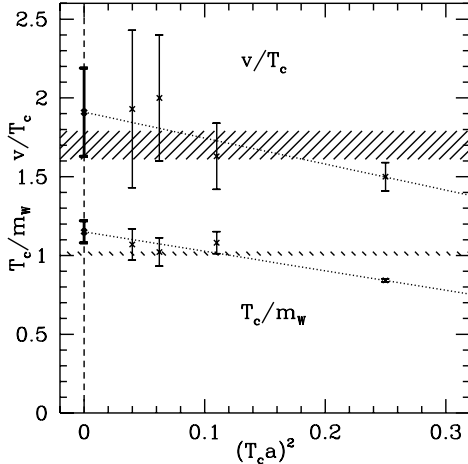}}
\caption[a]{The normalized jump and the critical temperature
in the continuum limit.}\label{jump}
\end{figure}

The bare squark mass parameters $m_Q^2,m_U^2,m_D^2$ receive quadratic 
renormalization corrections. As it is well known, one-loop lattice perturbation
theory is not sufficient to reliably determine these corrections, 
thus we use the following method. 
We first determine the position of the non-perturbative CB
phase transitions in the bare quantities (e.g.\ the triple point
  or the T=0 transition for $m_U^2$ in Fig.~\ref{phase}). These quantities are
compared with the prediction of the continuum perturbation theory,
which gives the renormalized mass parameters on the lattice.

\begin{figure}[t]
\centerline{\epsfxsize=6.9cm\epsfbox{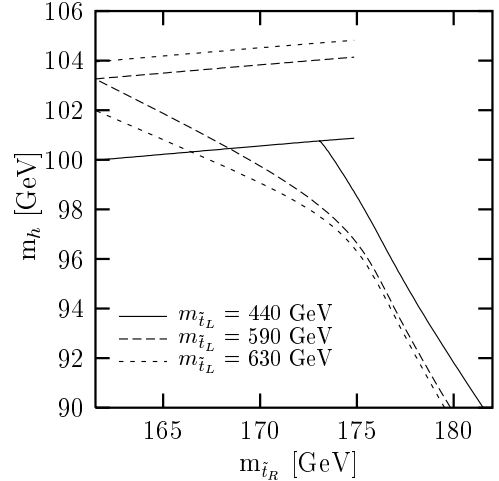}}
\caption[a]{The cosmologically relevant $v/T_c>1$ region is below the lines.
}\label{cosm}
\end{figure}

Fig.~\ref{jump} contains the continuum limit extrapolation for
the normalized jump of the order parameter ($v/T_c$: upper data)
and the critical temperature ($T_c/m_W $: lower data). The shaded
regions are the perturbative predictions at our reference point 
(see above) in the continuum. Their widths reflect 
the uncertainty of our reference point, which is dominated by the error of 
$m_h$. Note that $v/T_c$ is very sensitive to $m_h$, which results in the 
large uncertainties. Results obtained on the lattice  and in
perturbation theory agree reasonably within the estimated uncertainties.
(It might well be that the $L_t$=2 results are not in the scaling region;
leaving them out from the continuum extrapolation the agreement between 
the lattice and perturbative results is even better.) 

Based on this agreement we
use one-loop perturbation theory without HTE to determine cosmologically
allowed regions in the $m_{{\tilde t}_R}$  vs. $m_h$ plane of the full 
MSSM (including fermions, $m_A =500$ GeV), see Fig. \ref{cosm}. The two 
lines for each $m_{{\tilde t}_L}$ (which intersect for lower values of 
$m_{{\tilde t}_L}$) 
correspond to upper bounds resulting from  $v_n/T_c=1$  
(steeper curves, B1) and the T=0 maximum MSSM Higgs mass (B2). 
$v_n$ is the non-perturbative Higgs 
expectation value, assumed to be  larger than the perturbative one by 14\%, 
a correction factor obtained  in the bosonic model (cf. Fig. \ref{jump}). 
For large $m_Q$ (e.g. 600 GeV, $m_{{\tilde t}_L}$=630 GeV) the region below 
B1 is below B2. Decreasing
$m_Q$ B2 decreases and B1 increases. At $m_Q$=560 GeV 
($m_{{\tilde t}_L}$=590 GeV) B1 and B2 intersect at
the CB value of $m_{{\tilde t}_R}$. Since B2 is almost  constant this 
yields the overall maximum Higgs mass for a successful baryogenesis.
For even smaller $m_Q$ ($m_{{\tilde t}_L}$) both B1 and B2 are relevant. 
Note that the maximum 
Higgs mass corresponds to a finite value of $m_Q \approx$560 GeV,
yielding $m_h=103 \pm 4$ GeV 
(including also the uncertainties from Fig. \ref{jump} and the difference 
between the one and two-loop maximum Higgs mass calculations \cite{Carena}).

\begin{figure}[t]
\centerline{\epsfxsize=6.5cm\epsfbox{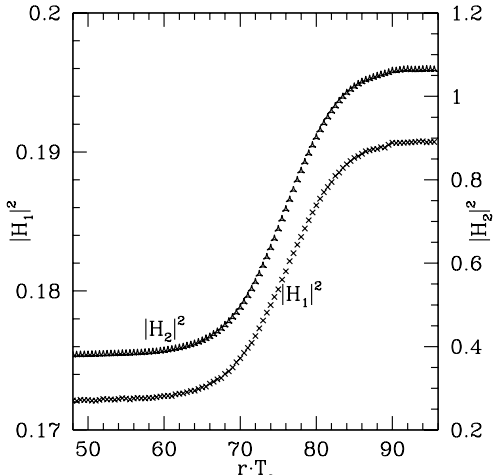}}
\caption[a]{The profile of the bubble wall for both of the Higgs fields 
for the lattice $2\cdot L^2 \dot 192$.}
\label{wall}
\end{figure}

In order to produce the observed baryon asymmetry, a strong first order
phase transition is not enough.
According to standard MSSM baryogenesis scenarios \cite{mssm_gen} the generated
baryon asymmetry is directly proportional to the variation
of $\beta$ through the bubble wall separating the Higgs and symmetric
phases.
By using elongated lattice ($2\cdot L^2 \cdot 192$), $L$=8,12,16 
at the transition point we study
the properties of the wall. In our simulation procedure
we restrict the length of one of the Higgs fields
to a small interval between its values in the bulk 
phases. As a consequence, the system fluctuates
around a configuration with two bulk phases and 
two walls between them. In order to have the smallest
possible free energy, the wall is perpendicular to the 
long direction. We eliminate the effect of the remaining
zero mode by shifting the wall of
each configuration to some fixed position. Fig.\ref{wall}~
gives the bubble wall profiles for both Higgs fields.
The measured width of the wall is 
[A+B$\cdot \log (aLT_c)]/T_c $ A=10.8$\pm$.1 and B=2.1$\pm$.1. 
This behavior indicates that the bubble wall is
rough and without a pinning force of finite size its width diverges
very slowly (logarithmically) \cite{Jasnow}.
For the same bosonic theory the perturbative approach predicts 
$(11.2\pm1.5)/T_c$ for the width. 

Transforming the data
of Fig.~\ref{wall} to $\vert H_2 \vert^2$ as a function of $\vert H_1 \vert^2$, 
we obtain $\Delta \beta=0.0061 
\pm 0.0003$.
The perturbative prediction at this point is $0.0046\pm0.0010$.
Thus perturbative studies such as \cite{wall_pert} 
are confirmed by non-perturbative results.

To summarize, we presented 4d lattice results on the EWPT in the MSSM.
Our simulations were carried out in the bosonic sector of the MSSM. We
found quite a good agreement between lattice results and our one-loop
perturbative predictions. Using this agreement together with a careful 
analysis of its uncertainties, we determined the upper
bound for the lightest Higgs mass for a successful baryogenesis in the
full (bosonic+fermionic) MSSM, which turned out to be  
($103 \pm 4$ GeV) consistent with the 3d analysis ($\approx 105$
GeV). We analyzed the bubble wall profile separating the Higgs and
symmetric phases. The width of the wall and the change in
$\beta$ is in fairly good agreement with perturbative predictions for 
typical bubble sizes.
Both the upper bound for $m_h$ and the smallness of $\Delta \beta$
indicate that experiments allow just a small window for MSSM
baryogenesis.

Details of the present analysis will be discussed in a forthcoming
publication \cite{mssm_prd}.

This work was partially supported by
Hungarian Science Foundation Grants 
OTKA-T22929-29803-M28413/FKFP-0128/1997.
The simulations were carried out on the
46G PC-farm at E\"otv\"os University.

\end{document}